# When is an action caused from within? Quantifying the causal chain leading to actions in simulated agents

Bjørn Erik Juel[1], Renzo Comolatti[2], Giulio Tononi[3] and Larissa Albantakis[3,*]
[1]Department of Molecular Medicine, Brain Signalling Group, University of Oslo, Oslo, Norway
[2]Institute of Science and Technology, Federal University of Sao Paulo, SP, Brazil
[3]Department of Psychiatry, Wisconsin Institute for Sleep and Consciousness, University of Wisconsin-Madison, WI, USA
*albantakis@wisc.edu

**Abstract**

An agent's actions can be influenced by external factors through the inputs it receives from the environment, as well as internal factors, such as memories or intrinsic preferences. The extent to which an agent's actions are "caused from within", as opposed to being externally driven, should depend on its sensor capacity as well as environmental demands for memory and context-dependent behavior. Here, we test this hypothesis using simulated agents ("animats"), equipped with small adaptive Markov Brains (MB) that evolve to solve a perceptual-categorization task under conditions varied with regards to the agents' sensor capacity and task difficulty. Using a novel formalism developed to identify and quantify the actual causes of occurrences ("what caused what?") in complex networks, we evaluate the direct causes of the animats' actions. In addition, we extend this framework to trace the causal chain ("causes of causes") leading to an animat's actions back in time, and compare the obtained spatio-temporal causal history across task conditions. We found that measures quantifying the extent to which an animat's actions are caused by internal factors (as opposed to being driven by the environment through its sensors) varied consistently with defining aspects of the task conditions they evolved to thrive in.

## Introduction

By definition, agents are open systems that can dynamically and informationally interact with their environments through sensors and actuators. However, identifying which particular set of events within or outside the agent caused it to act in a certain way is not straightforward, even if its internal structure and dynamics can be assessed in detail. This is demonstrated particularly well by the problem of accountability we currently face with respect to artificial intelligence (Doshi-Velez et al., 2017). While we can, in principle, record all network parameters of a (deep) neural network, such as AlphaGo (Silver et al., 2016), we still lack a principled set of tools to understand *why* the network performed a particular action (Metz, 2016).

Here, we address this issue using artificial agents ("animats") controlled by Markov Brains (MBs) (Hintze et al., 2017) as a model system of evolved agents, to which we apply a novel formalism for analyzing actual causation (AC) ("what caused what") in complex networks of interacting elements (Albantakis et al., 2019). Although there is no single widely accepted account of (actual) causation (Illari et al., 2011; Halpern, 2016), the AC framework presented by Albantakis et al. (2019) was specifically developed to identify and quantify the strength of the direct causes of any occurrence (subset of network nodes in a particular state at a particular time) within such systems. Notably, this formalism not only considers causes of single-variable occurrences, but also evaluates multivariate causal dependencies.

Given an appropriate model system of behaving agents, the AC framework may serve as a tool for assessing the actual causes of an agent's actions, by characterizing the actual causes of its motor actuators. For this purpose, animats controlled by MBs are particularly suited: MBs are a class of evolvable neural networks that receive sensor inputs and control motor outputs, and can be made small enough to allow for a complete causal and informational analysis, while remaining capable of evolving relatively complex behaviors (Edlund et al., 2011; Albantakis et al., 2014). As MBs may exhibit sparse, recurrent connectivity between their nodes ("neurons"), they resemble biological neural networks more closely than conventional machine-learning systems.

In this study, we demonstrate how the AC analysis can be utilized to evaluate the extent to which an animat's actions are "intrinsic" (caused by internal occurrences) rather than "extrinsic" (caused by sensor inputs, which are driven by the environment). To that end, we evolved animats to solve a perceptual-categorization task (Beer, 2003; Albantakis et al., 2014) under three task conditions with varying demands for memory and context-dependent behavior. As shown by Albantakis et al. (2014), animats evolved in more complex environments relative to their sensor capacity develop MBs with more densely connected nodes, more internal mechanisms, and higher integrated information, indicat-ing higher intrinsic causal complexity (Oizumi et al., 2014). Here, we exhaustively quantify the direct actual causes of the animats' actions, and expand the AC framework in order to trace back the causal chain ("causes of causes") leading up to a given action. In line with Albantakis et al. (2014), we hypothesize that the causal contribution of internal nodes to an animat's actions and its preceeding causal chain will be higher in animats evolved in more complex task conditions. Moreover, we expect that the causal chain will reflect task-specific demands for memory.

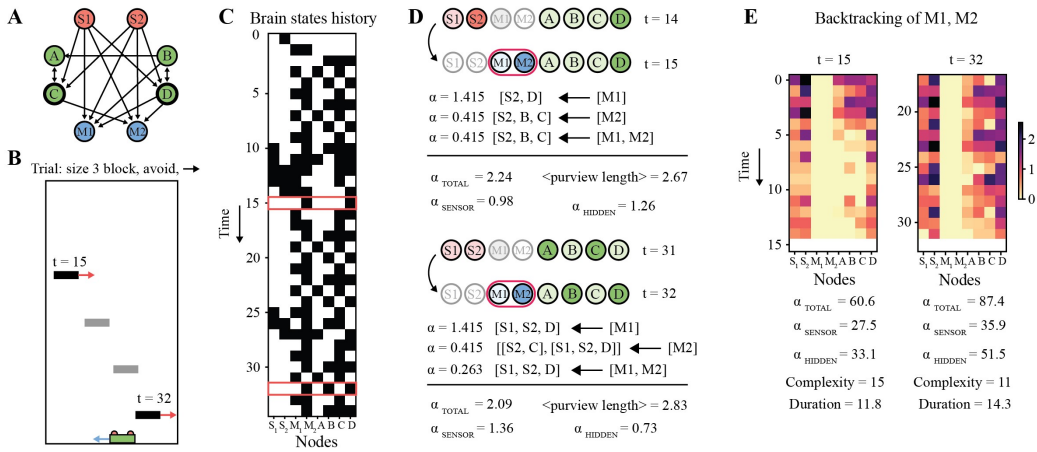

Figure 1: **Overview of the task environment, the animat, and analyses used in this study.** (A) An animat's Markov Brain with connections between sensors (red), motors (blue), and hidden nodes (green). Thick borders around nodes C and D indicate self connections. (B) Visualization of the simulated world with the animat at the bottom. (C) Time series of the animat's brain state during the trial. Black indicates that a node is 'on' ('1'), white means 'off' ('0'). The red frames highlight the two brain states ($t = 15$ and $t = 32$) whose direct causes and causal chains are shown in (D, E). (D) Direct causes of motor states in two example transitions. Dark color shades indicate that a node is 'on' ('1'), lighter shades mean 'off' ('0'). Listed below each transition are the direct actual causes (with corresponding causal strengths $\alpha$) for the states of M1, M2, and (M1, M2), together with the characteristic quantities used in the statistical analysis. (E) The causal chains obtained by tracing the actual causes of the motor state $(M1, M2) = 01$ ("move right") back in time for 15 time steps, starting at time $t = 15$ (left) and $t = 32$ (right). Each row in the pattern shows the summed causal strength ($\alpha$, indicated by color) that each node contributes to the direct causes of the previous time step. Listed below are the summary measures used to characterize the causal chains.

## Methods

To test our hypotheses, we utilize and extend a formal framework of actual causation to assess and trace back the causes of an agent's actions and apply it to artificial organisms ("animats") evolved *in silico* under several task conditions.

The artificial evolution experiments were conducted using the open source software package MABE (Modular Agent Based Evolver) (Bohm et al., 2017; Hintze et al., 2017). Software to identify and quantify direct actual causes is available as part of the PyPhi integrated information toolbox (Mayner et al., 2018). Finally, newly developed scripts that iteratively evaluate the actual causes of previously identified causes are available on GitHub (Comolatti and Juel, 2019).

## Data generation: Animat evolution and simulation

Evolution simulations were initialized using MABE's standard parameter settings with further agent and environment specifications described below. The agent types and task environments investigated in this study were adopted from Albantakis et al. (2014).

**Task environment.** Animats were evolved in the 'ComplexiPhi' world, a 35x16 unit grid with periodic boundary conditions, in which the animat has to move left or right to catch or avoid falling blocks of specific sizes (Figure 1A). Across trials, an animat (3 units wide) is placed at all positions along the bottom of the world, while blocks (one per trial) are positioned in the top left corner, falling to the left or right. The block is 'caught' if it overlaps with the animat when it reaches the bottom, otherwise it is 'avoided'.

**Markov Brains (MBs).** Animats are equipped with MBs. Each MB consists of up to 8 binary nodes: up to 2 sensors (S1 and S2), 2 motors (M1 and M2), and 4 hidden nodes (A, B, C, and D), whose function and connectivity is specified by hidden markov gates encoded in the animat's genome, as described in (Hintze et al., 2017). The global update function of the resulting neural network can be described by a state transition probability matrix (TPM). The TPM specifies the probability of an animat's MB transition-ing between any two states, thus completely describing its dynamics. Here, we specifically evolved animats with deterministic TPMs. Nevertheless, the causal analysis described below can be applied to probabilistic systems as long as they fulfill the causal Markov condition (Janzing et al., 2013; Albantakis et al., 2019).

During a trial, an animat's sensor is activated if a block is positioned directly above it at any height, and otherwise remains 'off' ('0'). The two sensors are positioned on each side of the animat, leaving a gap of 1 unit between them. The state of the remaining nodes updates according to the animat's TPM. However, motors are reset (set to 'off' ('0')) before each update, effectively excluding any feedback from the motors to the hidden nodes or sensors. The motor state determines the animat's action at each time step ('10': move left, '01': move right, '00' or '11': stand still). This process repeats until the block reaches the bottom of the world, at which point the success of the animat is recorded before the next trial begins.

**Genetic algorithm and task fitness.** Each evolution simulation is initiated with a population of 100 animats with random circular genomes. At each new generation, this pool of genomes is subject to fitness-based selection and mutation, which allows the animats to adapt to higher fitness across generations. An animat's fitness, $F$, is determined by its percentage of successful trials (correctly caught or avoided blocks). After each generation the genetic algorithm draws a new sample of 100 animats (with replacement) based on an exponential measure of $F$ (roulette wheel selection). Before the next generation, the genome of each selected animat mutates using point mutations, deletions, and duplications (Albantakis et al., 2014; Hintze et al., 2017). For each task condition, 50 populations of animats were initialized, and evolved independently for 30,000 generations.

**Task conditions.** Animats evolve under three conditions varying in difficulty relative to their sensor capacity:

- *Baseline (BL) condition*: catch blocks of size 1, avoid blocks of size 3, using 2 sensors;

- *One sensor (1S) condition*: catch blocks of size 1, avoid blocks of size 3, using only 1 (the left) sensor;

- *Hard task (HT) condition*: catch blocks of size 1 and 4, avoid blocks of size 2 and 3, using 2 sensors.

Thus, animats in the BL and 1S conditions perform the same task, while animats in BL and HT conditions use the same number of sensors. Compared to BL, in conditions 1S and HT additional computations across multiple time steps are necessary to distinguish which blocks have to be caught or avoided. Nevertheless, some internal memory is necessary in all conditions to identify whether the block is moving to the left or right.

**Data processing: Causal analysis**

**The Actual Causation (AC) framework.** Here, we briefly describe the relevant concepts of the AC framework by Albantakis et al. (2019). For details and formal definitions of the terminology, we refer to the original publication.

Given a transition $s_{t-1} \prec s_t$ between two subsequent states of a discrete dynamical system of interacting elements $S$, the AC formalism allows identifying the actual causes of occurrences at time $t$ from the set of occurrences at time $t-1$ based on a quantitative counterfactual analysis. Here, "occurrence" simply denotes a subset of network nodes in a particular state (for example a motor in state 'off': M1 = 0). In the AC framework, an occurrence $x_{t-1} \subseteq s_{t-1}$ may only be a cause of another occurrence $y_t \subseteq s_t$, if $y_t$ makes it more likely that $x_{t-1}$ has actually occurred. A "higher-order" occurrence (the joint state of a set of multiple nodes) may specify its own cause $x_{t-1}$, if it raises the probability of $x_{t-1}$ more than parts of the occurrence do when separated by a partition $\Psi$ (Oizumi et al., 2014; Albantakis et al., 2019). This difference in probabilities indicates the *causal strength* ($\alpha$) with which $y_t$ determines $x_{t-1}$. In simplified terms, $\alpha = \min_\Psi \left( \log_2 \frac{p(x_{t-1}|y_t)}{\Psi(p(x_{t-1}|y_t))} \right)$, where $\Psi$ partitions $p(x_{t-1} \mid y_t)$ into $p(x_{1,t-1} \mid y_{1,t}) \times p(x_{2,t-1} \mid y_{2,t}))$. $\alpha$ can be viewed as the irreducible information that an occurrence specifies about its cause. The *actual* cause of an occurrence $y_t$ is then defined as the subset $x^*_{t-1} \subseteq s_{t-1}$, for which $\alpha(x^*_{t-1}, y_t) = \alpha^{\max}(y_t)$. The set of nodes that constitutes the actual cause ($x^*_{t-1} \subseteq s_{t-1}$) is termed the *cause purview* of the occurrence in question, and the number of nodes in the purview is a measure of how distributed the cause is.

**Direct actual causes of motor states.** To identify the actual causes of an animat's action (M1 and M2 being in a particular state), and to quantify their causal strength, we consider transitions from all inputs to the motors (i.e., the sensors and hidden nodes) at time $t-1$ to the motors at time $t$: $(S1, S2, A, B, C, D)_{t-1} \prec (M1, M2)_t$. For a given state of the motors $(M1, M2)_t$, there can be a maximum of three distinct actual causes among all subsets of $(S1, S2, A, B, C, D)_{t-1}$: one for the state of M1, one for the state of M2, and one for the "higher-order" state $(M1, M2)_t$ (if $(M1, M2)_t$ is irreducible to its partition into M1 and M2).

As an example, consider the animat transition shown in Figure 1D from $t = 14$ to $t = 15$. Here, $(S2, D)_{14} = (1, 1)$ is the actual cause of $M1_{15} = 0$ with $\alpha = 1.415$ bit. $(S2, B, C)_{14} = (1, 0, 0)$ is the actual cause of $M2_{15} = 1$ with causal strength $\alpha = 0.415$ bit. In addition, $(M1, M2)_{15} = (0, 1)$ also has its own actual cause $(S2, B, C)_{14} = (1, 0, 0)$ with $\alpha = 0.415$ bit. This means that $(M1, M2)_{15} = (0, 1)$ specifies an additional 0.415 bit of information about the state of $(S2, B, C)$ at $t = 14$ compared to $M1_{15} = 0$ and $M2_{15} = 1$ taken individually. Also, note that the cause purviews of the three occurrences contain different sets of nodes. For example, hidden node B is part of the actual cause of M2 in this transition, but is not in the cause purview of $M1_{15} = 0$. Moreover, there is no need for a node to be 'on' ('1') to be part of the cause of an occurrence. Finally, S1, for example, is not part of any actual cause even though it has a direct connection to both M1 and M2. In other words, $S1_{14} = 0$ does not contribute to "bringing about" the state of the motors at $t = 15$ in this particular transition. However, in the transition from $t = 31$ to $t = 32$, $S1_{31} = 0$ does contribute to the causes of $(M1, M2)_{32}$.

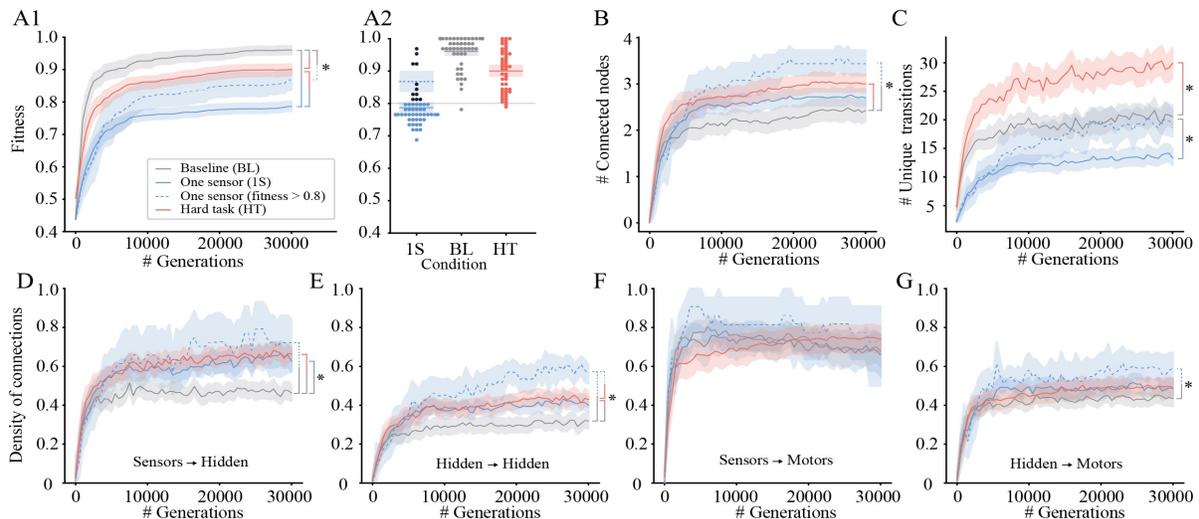

Figure 2: **Fitness, structural, and functional properties of animats.** (A1) Fitness of animats (with final fitness distributions in (A2)), quantified as the ratio of successful trials. (B) #connected nodes measures the number of hidden nodes with incoming and outgoing connections (maximally 4, A-D). (C) #Unique transitions across all trials within a generation estimates the dynamical complexity of the animats. (D)-(G) Density of connections between different node types quantified as the number of connections between the nodes divided by the total possible number of such connections. The lines show the bootstrap resampled means for the population (the shaded area indicates the 95% CI of the mean). Statistically significant difference between conditions are indicated by the colored bars (and stars). The legend in (A) indicates the coloring for all panels.

Thus, for every animat, we find the direct actual causes of the motor state in every unique transition and calculate several measures to quantify the degree to which the motor state was caused from within (see legend of Figure 3).

**Quantifying the causal chain: the backtracking analysis.** We also perform a 'backtracking analysis' of the causes of the motor states, for all transitions in a trial past $t = 15$ (Figure 1E). This analysis amounts to identifying and characterizing the chain of causes leading up to an action.

Having quantified the direct actual causes of a transition $x_{t-1} \prec y_t$ (where $x_{t-1}$ is the state of the sensors and hidden nodes at time $t-1$ and $y_t$ is the state of the motors at time $t$), we define a joint purview $z_{t-1} \subseteq x_{t-1}$ as the union of all identified cause purviews. In other words, $z_{t-1}$ is the state at time $t-1$ of all elements contributing to the actual causes of the motor occurrences, $M1_t$, $M2_t$, and $(M1, M2)_t$. In the example shown in Figure 1D (top), this corresponds to $z_{14} = \{(S2, B, C, D)_{14} = (0, 1, 0, 1)\}$.

Iteratively, we then proceeded to identify and quantify the actual causes for the transitions $x_{t-2} \prec z_{t-1}$, $x_{t-3} \prec z_{t-2}$, etc., thus tracing back the causal chain of the observed motor state at time $t$. This process is repeated until all cause purviews in the causal account contain only sensors (indicating that the cause is completely "extrinsic") or upon reaching $t - 15$ (to make results comparable across time steps).

For every animat, we find the causal chain leading to each motor state (after $t = 15$) and calculate several measures aimed to quantify the "intrinsicality" of the backtracking pattern (see legend of Figure 4).

### Statistics

Throughout this work, we use bootstrap resampling to estimate means and to calculate 95% confidence intervals between the 2.5th and the 97.5th percentile of the resampled distribution ($CI_{95\%} = [P_{2.5\%}, P_{97.5\%}]$). Although data samples generating overlapping confidence intervals may still differ significantly, we take non-overlapping confidence intervals as an indicator of a statistically significant difference between conditions.

## Results

Our simulated evolution experiments under three task conditions reproduce earlier findings reported in (Albantakis et al., 2014). Average fitness is higher in the baseline (BL) condition, than in both the one-sensor (1S) and the hard-task (HT) conditions (Figure 2A). In addition, the fitness among animats in the HT condition is significantly higher than in the 1S condition, indicating a hierarchy of overall difficulty among the three conditions: BL < HT < 1S. Since fewer animats achieved a final fitness of at least 80% by generation 30,000 for the 1S than the BL and HT conditions (11/50 for 1S vs 49/50 and 48/50 for BL and HT), we use the subset with fitness > 80% as the default population in condition 1S (representing successful evolution), unless otherwise stated.

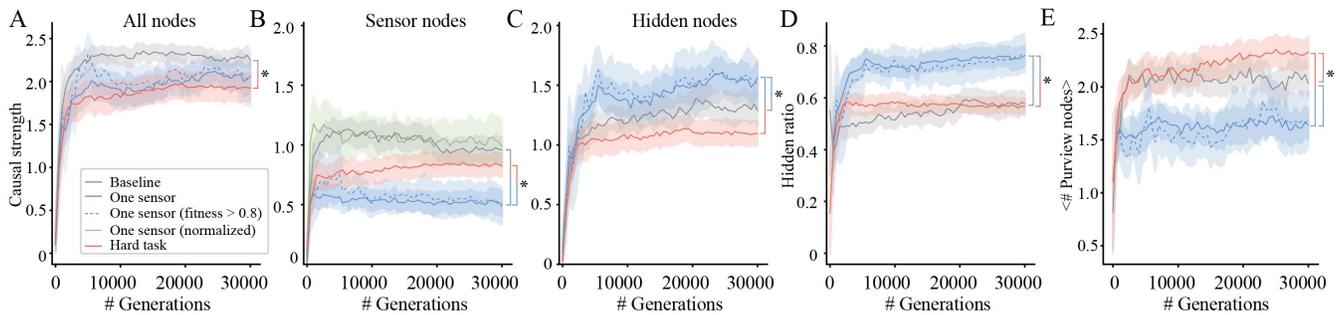

Figure 3: **Analysis of direct causes for actions as a function of generation in different task conditions.** (A) Total causal strength: summed $\alpha$ across the identified direct causes of M1, M2, and (M1, M2). (B) Sensor causal strength: summed $\alpha$ of the sensor portion in the cause purviews. (C) Hidden causal strength: summed $\alpha$ of the hidden node portion in the cause purviews. (D) Hidden ratio: the ratio of hidden and total causal strength per purview. (B-D) To compute the sensor and hidden portion of the causal strength, we simply multiplied the fraction of sensor and hidden nodes in each actual cause purview by its $\alpha$ value. We did not recompute $\alpha$ values for sensor or hidden node subsets, as these subsets do not correspond to actual causes themselves. (E) Total number of nodes in the cause purviews. All measures are averaged across the unique transitions across times and trials for each animat. Mean, 95% CI, and statistical significance are as in Figure 2.

**Structural and dynamic analysis of animats**

Generally, measures of the structural and dynamical complexity of the animats increase with fitness throughout evolution (Figure 2; the only exception is a decrease in connections from sensors to motors for 1S and BL in panel F).

In terms of the dynamical complexity exhibited by the animats, those evolved in the HT condition show the largest repertoire of unique state transitions (Figure 2C). Although the number of potential state transitions is smaller in condition 1S due to the reduced number of available nodes (7 vs. 8), the fittest 1S animats still compare to the BL condition.

Structurally, differences from the BL condition are most pronounced in the number of connected nodes (Figure 2B) and the density of connections to the hidden nodes (Figure 2D,E). In particular, the density of connections between hidden nodes reflects the hierarchy of task difficulty. On the other hand, differences in connections to the motors are smaller (Figure 2D,G).

**Actual Causation analysis: direct causes of actions**

To characterize the causes of the animats' actions across the three task conditions, we first analyzed the direct causes of their motor states for all unique transitions per animat. Here, only nodes directly connected to the motors may appear in the actual cause purviews of a motor occurrence (see Figure 2D,G). However, whether any particular input node contributes to the cause purviews may vary depending on the transition (see Figure 1D). Nevertheless, as shown in Figure 3, the differences in direct motor causes between conditions do not simply follow the pattern observed for the structural and dynamical analysis (Figure 2).

Animats in the HT condition show a lower total and hidden causal strength (Figure 3A and C), but a higher number of nodes in the cause purviews (Figure 3E) than animats from the BL condition. The 1S condition exhibits lower sensor causal strength, but higher hidden causal strength than conditions BL and HT, and correspondingly, also a significantly higher hidden ratio (Figure 3C-E). Furthermore, the number of purview nodes is significantly lower in the 1S condition than in BL and HT (Figure 3E).

Thus, although there are no clear differences in the density of connections to motors between the conditions, the direct cause analysis seems to distinguish the more difficult conditions from the BL condition by the number of nodes in the purview and the way the causal strength is distributed among different types of nodes, albeit in opposite directions.

**Actual Causation analysis: Backtracking analysis.**

Next, we investigated whether considering the causal chain ("causes of causes") leading to an animat's action yields additional information about the causal structure of the animat and the causes of its actions. Here, every node with a directed path towards a motor may contribute to the causal chain given sufficient backsteps.

In contrast to the direct cause analysis, in the backtracking analysis both the 1S and HT conditions differ significantly from BL in a consistent manner on several measures. First of all, the total, sensor, and hidden causal strengths (Figure 4A-C) of the backtracking patterns are all higher (for the fittest animats) in 1S and HT than in BL. However, only the 1S condition has a higher hidden ratio than BL (Figure 4D), as in the direct cause analysis (Figure 3D). Thus, animats in the HT condition seem to require more involvement from both sensors and hidden nodes than the BL condition to successfully complete their task. On the other hand, animats in the 1S condition seem to compensate their missing sensor by relying more on hidden nodes, resulting in a higher hidden ratio compared to the BL condition. In addition, both the complexity (Figure 4E) and duration (Figure 4F) of the backtracking patterns were higher in the 1S and HT conditions than in the BL condition, reflecting the higher memory requirements in these two conditions.

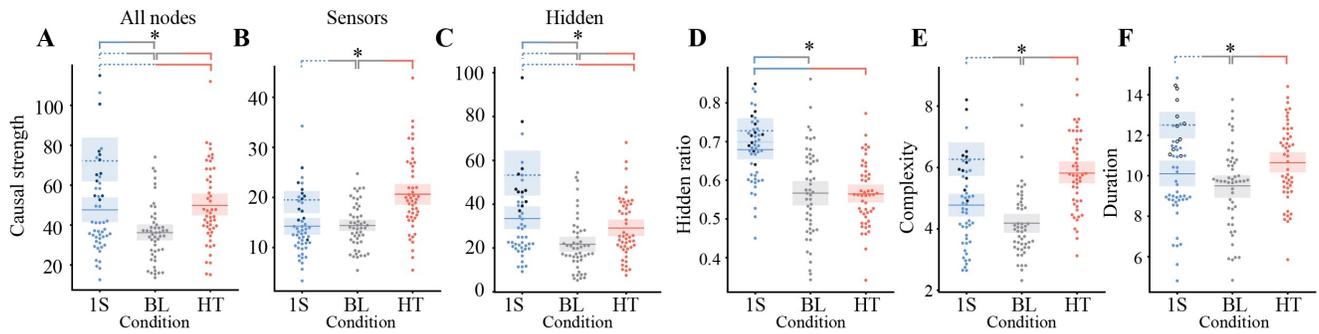

Figure 4: **Results of backtracking analysis from animats in the last generation.** (A) Total causal strength: summed $\alpha$ across the backtracking pattern. (B) Sensor causal strength: summed $\alpha$ of the sensor portion in the backtracking pattern. (C) Hidden causal strength: summed $\alpha$ of the hidden node portion in the backtracking pattern. (D) Hidden ratio: the ratio of hidden and total causal strength. (E) Complexity of the causal chain: number of unique rows in a backtracking pattern. (F) 'Duration' measures the average length of the causal chain by calculating the area of the backtracking pattern and normalizing by the max number of sensor and hidden nodes in the animat (6 for the BL and HT conditions and 5 for the 1S condition). Each dot corresponds to the value for one animat (averaged over times and trials) in the last generation of the evolution. Colors are as in previous figures, with the dark blue dots marking animats from the 1S condition with fitness above 0.8. Shaded patches indicate 95% confidence intervals. Statistically significant differences between populations are indicated by the colored bars (and stars).

Finally, the systematic effect of including only the fittest animats from the 1S condition was apparent across all evaluated characteristics of the backtracking patterns, and the hierarchy of task difficulty across the three conditions (BL < HT < 1S) observed in the structural analysis (Figure 2) reemerged for the backtracking analysis.

## Discussion

In this paper we applied measures of structural and dynamical complexity, as well as a novel actual causation (AC) framework (Albantakis et al., 2019), to characterize the actions of simulated agents that evolved to solve perceptual-categorization tasks. With the structural and dynamical analysis, we confirmed findings from previous studies indicating that more demanding task conditions lead to the evolution of animats with more interconnected 'brains' with a higher capacity for computations and memory (Albantakis et al., 2014). Using the AC framework, we identified and quantified the direct causes of the animat's actions as well as their preceeding causal chains. To assess the degree to which an animat's actions were "caused from within" (as opposed to being externally driven through its sensors), we moreover quantified the relative contribution of its hidden and sensor nodes to the cause purviews and backtracking pattern of its motor states.

As discussed in more detail below, we found that the different types of analyses revealed different aspects about the animats' causal structure and the task conditions under which they evolved. While some measures reflected the hierarchy of task difficulty across the three conditions (BL < HT < 1S), the direct actual causes in particular highlighted differences between 1S and HT.

As hypothesized, the causal chains leading to actions in animats evolved in difficult conditions (hard tasks relative to sensor capacities) with higher demands for internal memory were characterized by higher total causal strength, and reverberated longer within the animat itself than in animats evolved in the simpler baseline condition.

In all, our results suggested that the "intrinsicality" of the direct causes and the causal chain preceding an agent's actions may serve as a useful indicator of its intrinsic complexity and degree of causal autonomy (see also Marshall et al. (2017); Bertschinger et al. (2008)), while the number of nodes constituting the cause purviews, as well as the complexity and duration of the causal chains may reflect its context-sensitivity.

**Differences between conditions**

Both the 1S and HT conditions require more internal memory to distinguish which blocks have to be caught or avoided compared to the BL condition. This task property is reflected in the higher number of connected nodes (Figure 2B) and the measures assessed in the backtracking analysis (Figure 4), and also underlies the observed hierarchy in task difficulty (Figure 2A).

However, in the HT condition, more blocks need to be classified than in BL and 1S. Nevertheless, the animats can only make use of the same repertoire of possible actions. Which action is chosen is thus more context-dependent in condition HT than in the other two conditions. The higher context-dependency of condition HT may explain the opposing results between HT and 1S in the direct causes of the animats' actions (Figure 3): the direct cause purviews in the HT condition are larger and more distributed across sensors and hidden nodes, leading to a similar hidden ratio as BL. In contrast, animats in the 1S condition have a higher hidden ratio due to increased memory requirements, but less distributed computation. In sum, our results suggest that the backtracking analysis measures are mostly affected by aspects related to memory requirements, while the direct cause analysis capture context-sensitivity and distributed computation.

**Causal analysis**

Given that the results of the causal analysis, at least in part, reflect differences in the structural and dynamical properties of the animats, the advantages of a computationally demanding causal analysis may not be immediately clear. For example, one could argue that observed differences between conditions in the causal analysis might be explained by structural properties of the animat populations (such as the longer causal chains with higher hidden causal strength being explained by more, and more densely interconnected, nodes). However, there are at least two reasons for applying the AC analysis in addition to more standard approaches.

First, the AC analysis specifically takes an animat's mechanistic, counterfactual structure into account (see also Shalizi et al. (2005)). Therefore, it may describe aspects of the system that cannot be captured by purely structural, or dynamical, informational, or correlational measures based on observed data only (Marshall et al., 2017). For example, we hardly found significant differences between task conditions regarding the inputs to the motor nodes (Figure 2F,G). Yet, the sensor and hidden causal strength varied significantly across conditions (Figure 3B,C and Figure 4B,C).

Secondly, the AC analysis is applied to each individual transition independently and can identify the causes and intrinsicality of each specific action (motor state), giving a state-dependent description of the animat behavior (see also Lizier et al. (2014) and Beer and Williams (2014)). As can be seen from the example in Figure 1, the same action being performed by the same animat at two different times may have distinct causes depending on the past states of the rest of the animat. Correlations between an agent's structural/dynamical properties and the results of the causal analysis may thus only become apparent when averaging across many transitions as done here. In future work, it could be investigated how the actual causes of an animat's actions change on a trial-by-trial basis, for specific block sizes, the direction of motion, or whether a block should be caught or avoided. State-independent measures that characterize the animat as a whole cannot assess such questions, but may still serve as useful indicators for a system's capacity for internally caused motor states.

Of course, alternative formalisms for measuring actual causes and causal chains exist (e.g. Datta et al. (2016); Weslake (2015)), which might also be applicable to artificial agents. Nevertheless, the AC framework used here was specifically developed for discrete dynamical systems of interacting elements, such as Markov Brains, which makes it particularly suited for the present study. An interesting question is under which circumstances causal measures effectively exceed dynamical or information-theoretical approaches in elucidating an agent's behavior (e.g., Beer and Williams (2014); Lizier et al. (2014)).

Finally, we did not directly consider issues regarding causal transitivity in this work. The question of whether (and when) the "causes of causes" of an occurrence are themselves causes, is still highly debated. To answer such questions, our proposed approached must be further refined and adjusted accordingly.

**Towards a principled definition of agency**

On a more philosophical note, the definition of terms used here (such as agent and agency) should be revisited and clarified in future work. For example, throughout this paper we have been using the term agent to refer to the predefined set of nodes that comprise the animats under investigation. And if we define an agent loosely as some system that can sense and interact with its environment, this may not seem problematic. However, if we aim to understand agency more fundamentally, we would also need a way to determine which subset of nodes within a larger set of elements actually constitutes the agent. For this we require a more stringent definition of an agent, as we cannot assume that the borders of the agent itself can always be drawn *a priori*.

One example of such a more stringent definition could be that an agent is (1) an open physical system with stable, self-defined and self-maintained causal borders, with (2) the capacity to perform actions that causally originate within the system itself (Albantakis, 2018) (see also Polani et al. (2016) for an information-based alternative). In this context, it has been shown that the same causal principles on which the actual causation framework is based (Oizumi et al., 2014) can also be used to identify the causal borders of highly inte-grated subsets of nodes within larger networks, indicating that this type of causal analysis may be used to find autonomous systems that fulfill the first criterion for agency listed above (Marshall et al., 2017) (alternatively, see Friston (2013), Beer (2015), or Kolchinsky and Wolpert (2018)). In summary, we may draw upon and adhere to the theoretical structure and mathematical formalism used in the integrated information theory (IIT) of consciousness (Oizumi et al., 2014) to evaluate both parts of the proposed two-fold definition of agency. Thus, it seems possible that the IIT and AC formalism, taken together, may be used to relate concepts of agency, autonomy, causality, and consciousness within a self-consistent and principled theoretical framework.

## Acknowledgements

B.E.J received internationalization support from UiO:Life Science, the University of Oslo, and salary from European Unions Horizon 2020 research and innovation programme under grant agreement 7202070 (Human Brain Project (HBP)). L.A. receives funding from the Templeton World Charities Foundation (Grant #TWCF0196).